# Shell-Core Structural Anisotropy in Starch Granules Probed by Polarization Third-Harmonic Generation Microscopy

**Maria Kefalogianni [1,2], Leonidas Mouchliadis [2], Emmanuel Stratakis [2,*], and Sotiris Psilodimitrakopoulos [2,*]**

[1] Department of Physics, University of Crete, Voutes University Campus, GR-70013 Heraklion, Greece
[2] Institute of Electronic Structure and Laser, Foundation for Research and Technology-Hellas, 100 N. Plastira str., GR-70013 Heraklion, Greece
* Correspondence: sopsilo@iesl.forth.gr; stratak@iesl.forth.gr

**Abstract**

Lately the non-linear optical third harmonic generation (THG) microscopy is starting to emerge as a laboratory standard for label-free studies in biological samples. In this study, the THG signals produced from corn starch granules are investigated. In particular, the polarization-dependent THG (P-THG) signals emerging from the outer layer (shell) of the starch granules are compared with the P-THG signals originating from their inner portion (core). By rotating the linear polarization of the excitation beam, two distinct P-THG modulation patterns are revealed within single granules, corresponding to their shells and to their structurally different cores. These patterns are analyzed using a theoretical framework, that describes THG from an orthorhombic crystal symmetry, characteristic of corn starch. This allows us to extract point-by-point in the granules the ratios of the $\chi^{(3)}$ susceptibility tensor elements and the average molecular orientations. Then, the anisotropy-ratio (AR=$\chi^{(3)}_{xxxx}/\chi^{(3)}_{yyyy}$) is defined and used as a quantitative descriptor of the local molecular arrangements. Our results show that the shells and cores exhibit distinct AR values, probing the anisotropy in the molecular arrangements between the two regions. This study establishes P-THG as a powerful contrast mechanism for probing structural anisotropy in biological samples beyond conventional THG intensity-only microscopy.

**Keywords:** starch granules; non-linear microscopy; polarization third harmonic generation; orthorhombic crystal symmetry; structural anisotropy

## 1. Introduction

Starch is one of the most important polysaccharides and a basic constituent of our everyday diet. It is found in potatoes, wheat, rice and corn, and it is laid down in the form of granules. It consists of two types of molecules, amylose and amylopectin. The extensive degree of crystallinity in starch granules originates from the amylopectin molecules which are oriented radially with their non-reducing chain ends pointing towards the outer surface of the granule. This spherocrystalline assembly when observed under polarized light, shows a Maltese cross [1]. Furthermore, using a λ/4 waveplate, polarized light shows the positive birefringence which theoretically indicates radial orientation of the principal axis of the crystallites [2]. The crystalline structure of starch varies by botanical origin and falls into three general types: A-type, B-type, and C-type. Corn starches, belong to the orthorhombic A-type crystal structure [3,4]. Starch molecules are organized in alternating concentric crystalline and amorphous domains that form growth rings. The crystalline layer consists of ordered regions composed of double helices formed by short amylopectin branches, most of which are further ordered into structures known as the crystalline lamellae [5]. The outer layer of starch granules (shell) seems to differ from the inner portion (core), although the chemical composition is basically the same. The remnant of this envelope structure, commonly called a “ghost”, has been found to comprise primarily amylopectin (as determined by iodine staining) [6,7]. Techniques like electron microscopy [8], X-ray scattering [9] and nuclear magnetic resonance [10] are used to characterize the structure of starch granules. Optical approaches, like second-harmonic generation (SHG) microscopy, also offer valuable information on the structure of starch granules [11-14]. Lately, non-linear optical third harmonic generation (THG) microscopy is used to image starch [15-17]. In THG, three photons are combined to produce one of triple energy. THG signals are coherent with well-defined polarizations. The power of the THG signals depends on the state of the excitation polarization and the structure of the interacted samples.

In the present study polarization-dependent THG (P-THG) microscopy is performed in corn starch granules. By rotating the direction of the excitation linear polarization, series of images of the granules are acquired, for different directions of polarization. In these P-THG images, the THG signals’ intensity is modulating with the rotation of the polarization, giving rise to characteristic P-THG patterns. Importantly, the THG signals originating from the shell of the granule, modulate in a different manner than those originating from the core of the granule. After a point-by-point analysis, distinct P-THG modulation profiles are revealed. These emerge from different regions of the corn starch granules (i.e. the shell and the core). A theoretical framework is introduced for interpreting P-THG signals originating from an orthorhombic crystal symmetry, characteristic of corn starches. The ratios of the $\chi^{(3)}$ susceptibility tensor elements are extracted, from fits to the P-THG intensity profiles. An anisotropy ratio AR=$\chi^{(3)}_{xxxx}/\chi^{(3)}_{yyyy}$ is introduced and used as a metric of the molecular arrangement in the granules. Our analysis reveals distinct AR values for the cores and shells, reflecting structural anisotropy on molecular arrangements within single granules.

## 2. Materials and Methods

### *2.1. Sample Preparation*

The corn starch was purchased from the local food store. It was embedded with distilled water between two 25mm diameter, No.0

thickness, microscope cover glasses (Menzel Gläser), sealed with grease. The addition of water has significant effect on starch structure. Hydration self organizes amylopectin molecules in a more uniform spherical granule shape. Specifically, water allows the amylopectin helices to organize side by side in a smectic A-type structure [18].

### *2.2. Polarization-dependent THG microscope*

The experimental setup of our custom-build non-linear microscope is schematically shown in Figure 1 [19]. The excitation source is a femtosecond oscillator (FLINT, Light Conversion) centered at 1030nm, with repetition rate 80MHz and pulse duration 50fs (manufacturer specifications). The output beam is directed into an optical parametric oscillator (OPO) (Levante IR, APE) tuned to emit at 1542nm for THG excitation. The linear polarization state of the excitation beam is controlled by a half-wave retardation plate (AHWP10M-1600, ThorLabs) placed on a motorized rotational stage (8MRU, Standa). Raster-scanning of the beam is performed using a pair of galvanometric mirrors (6215H, Cambridge Technology). Using a pair of achromatic doublet lenses (forming a telescope), the beam emerging from the galvanometric mirrors is expanded approximately 7 times in order to fill the back aperture of the objective lens. Exiting the telescope, the beam is reflected on a mirror, at 45° (PFR10-P01, ThorLabs), placed at the motorized turret box of the microscope, just below the objective (Plan-Apochromat 40x/1.3NA, Carl Zeiss).

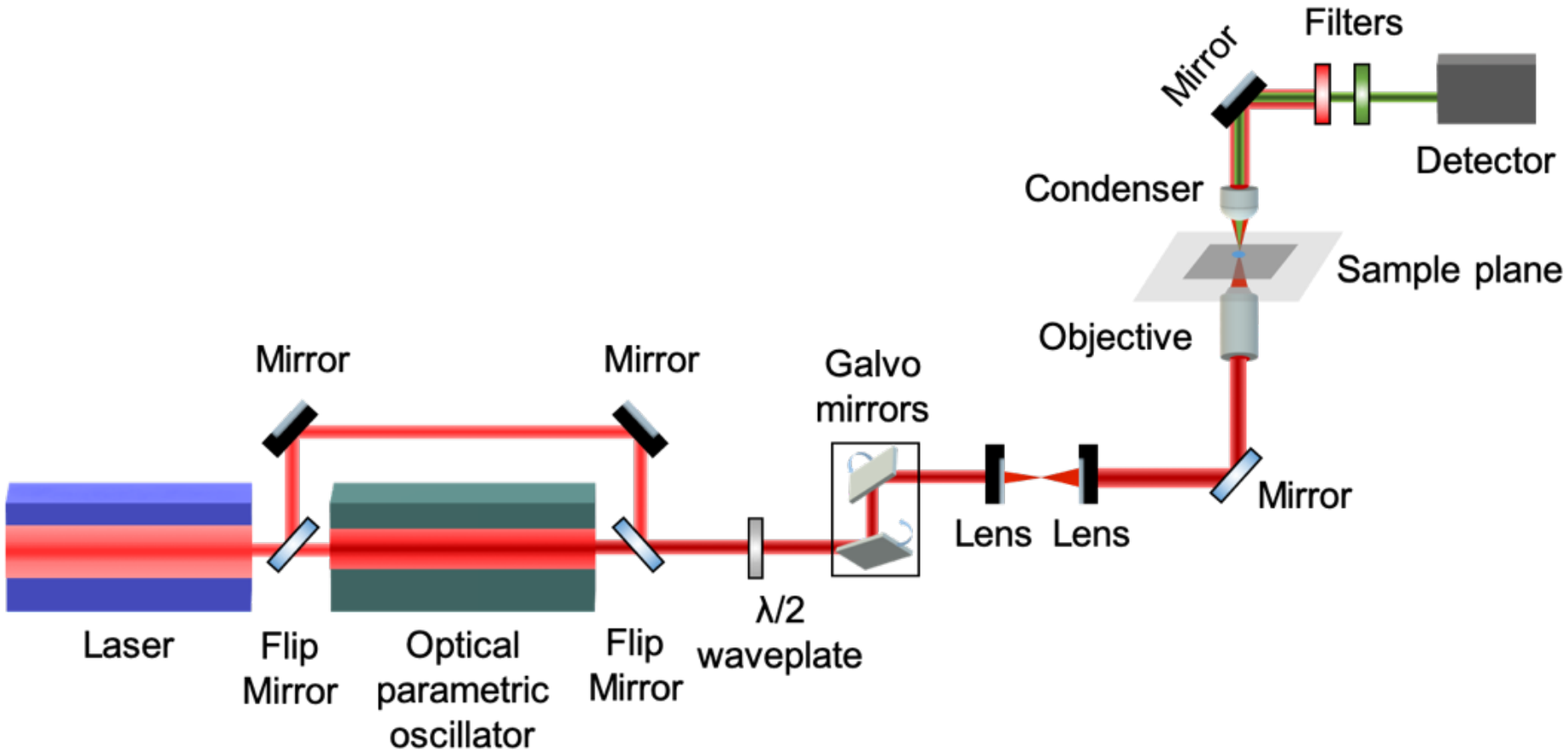

**Figure 1.** Schematic representation of the experimental setup for P-THG imaging. It is based on a fs laser beam coupled to a microscope. The fundamental, 1030 nm, pulses are converted to 1542 nm, by means of an optical parametric oscillator. Starch granules, excited by this wavelength, generate THG centered at 514 nm. The THG signals are recorded for different directions of the linear polarization of the excitation beam, via a rotating a half-wave plate. A pair of galvanometric mirrors enables raster-scanning of the *fs* beam, obtaining P-THG images of the starch granules, for every direction of polarization.

The THG (at 514 nm wavelength) is collected in the forward direction with a high-numerical aperture condenser lens (achromatic - aplanatic, 1.4NA, Zeiss) and is detected, after passing through a short-pass filter (FES0800, ThorLabs) and a band-pass filter (FF01-514/3, Semrock), by a photomultiplier tube (H9305-04, Hamamatsu). The galvanometric mirrors and the PMT are connected to a connector block (BNC-2110, National Instruments), which is interfaced to a PC through a DAQ (PCI 6259, National Instruments). Coordination of PMT recordings with the galvanometric mirrors for the image formation, as well as the movements of all the microscope motors, is carried out using LabView (National Instruments) software. One image of 500x500 pixels (measurements) is recorded in approximately 1s.

*2.3. Theoretical model of polarization-dependent THG in starch*

To model P-THG in starch granules two coordinate systems are introduced (Figure 2): the laboratory frame (X-Y-Z), and the sample frame (x-y-z). The laser beam propagates along the laboratory Z-axis, and its electric field is linearly polarized in the X–Y plane. The polarization angle, denoted by α, is defined with respect to the X-axis and is varied systematically during THG intensity measurements.

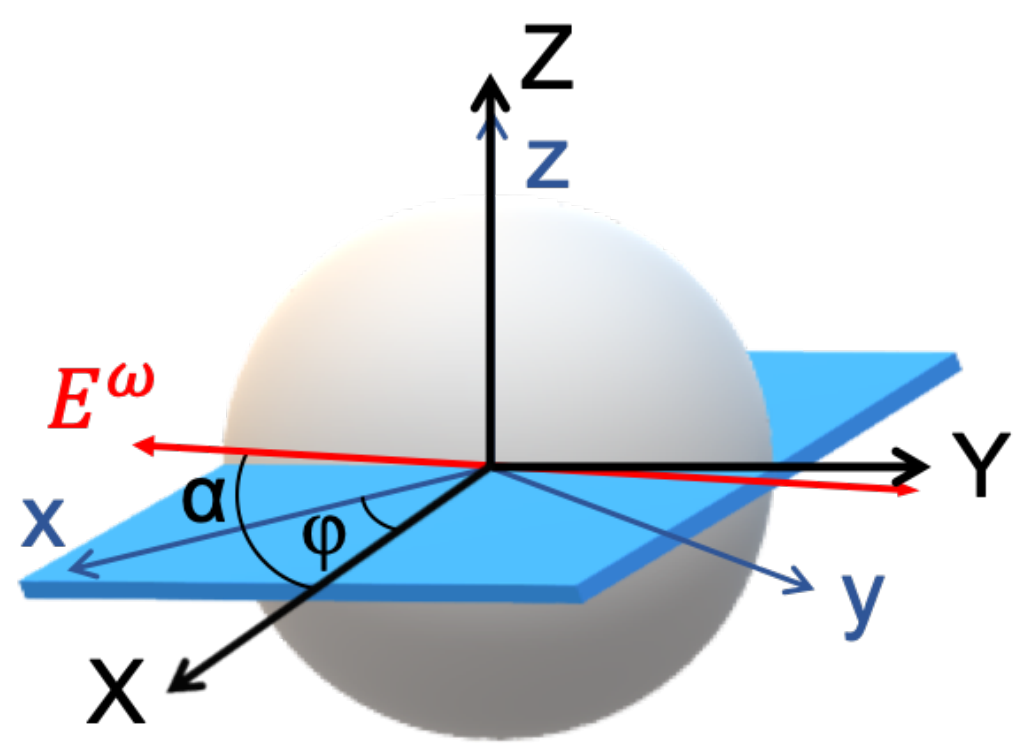

**Figure 2.** Coordinate systems used in the theoretical description of polarization-dependent THG in starch granules. The schematic illustrates the relationship between the laboratory coordinate system (X–Y–Z) and the sample coordinate system (x–y–z). At the equatorial plane of the granule, where the molecular axes lie predominantly in the x–y plane, the sample and laboratory coordinate planes (x–y and X–Y, respectively) are considered parallel (z=Z). The excitation electric field $E^{\omega}$ is linearly polarized and rotates within the X–Y plane at an angle α relative to the X-axis. The orientation of the sample is defined by the angle φ, representing the rotation of the sample x-axis relative to the lab X-axis. This geometric configuration enables modeling of the P-THG signals in a system with radial molecular organization like in the starch granules.

Based on polarization-dependent SHG studies, starch granules exhibit radial molecular organization in both the shell and the core [11]. An equatorial plane of the granule, where the long axes of the starch molecules predominantly lie in the sample's x–y plane (because of the radial organization), is considered. At this equator, the sample and laboratory planes are aligned such that z=Z, and the x–y plane of the sample is coplanar with the X–Y plane of the laboratory frame. The orientation of the sample frame with respect to the laboratory frame is defined by an in-plane rotation angle φ between the sample x-axis and the laboratory X-axis. In the sample frame, the incident electric field components are expressed as:

$$
\begin{aligned}
E_{\mathrm{x}}^{\omega} &= E_0 \cos(a-\varphi) \\
E_{\mathrm{y}}^{\omega} &= E_0 \sin(a-\varphi) \\
E_{\mathrm{z}}^{\omega} &= 0,
\end{aligned} \tag{1}
$$

where $E_0$ is the amplitude of the incident electric field. Effects such as birefringence and diattenuation due to scattering [20], and axial field components due to tight focusing [21] are neglected in this model for simplicity.

The induced third-order nonlinear polarization ($P^{3\omega}$) in the sample is governed by a third-order susceptibility tensor $\chi_{ijkl}^{(3)}$ [22]. For convenience, the contracted notation $\chi_{\mathrm{im}}^{(3)}$ is adopted where the index i∈{1,2,3} corresponds to x, y z , respectively, and the multi-index *jkl* is mapped to a single index *m* using standard nonlinear optics conventions [22]:

| *jkl* | *xxx* | *yyy* | *zzz* | *yzz* | *yyz* | *xzz* | *xxz* | *xyy* | *xxy* | *xyz* |
|---|---|---|---|---|---|---|---|---|---|---|
| *m* | 1 | 2 | 3 | 4 | 5 | 6 | 7 | 8 | 9 | 0 |

Because corn starch granules are SHG active (Figure S1), the orthorhombic crystal symmetry (*mm2*) that allows SHG [22] is chosen. Assuming the orthorhombic (*mm2*), as appropriate for corn starch [3,4], only a subset of the $\chi_{im}^{(3)}$ tensor components are non-zero. Specifically, the following nine elements contribute to the third-order response [19]: $\chi_{11}^{(3)}, \chi_{16}^{(3)}, \chi_{18}^{(3)}, \chi_{22}^{(3)}, \chi_{24}^{(3)}, \chi_{29}^{(3)}, \chi_{33}^{(3)}, \chi_{35}^{(3)}, \chi_{37}^{(3)}$. With these non-zero elements and the field components defined above, the third-order polarization $P^{3\omega}$ in the sample can be explicitly calculated as a function of the input polarization angle α and the sample orientation φ:

$$
\begin{pmatrix} \mathrm{P}_{\mathrm{x}}^{3\omega} \\ \mathrm{P}_{\mathrm{y}}^{3\omega} \\ \mathrm{P}_{\mathrm{z}}^{3\omega} \end{pmatrix} = \varepsilon_o \begin{pmatrix} \chi_{11}^{(3)} & 0 & 0 & 0 & 0 & \chi_{16}^{(3)} & 0 & \chi_{18}^{(3)} & 0 & 0 \\ 0 & \chi_{22}^{(3)} & 0 & \chi_{24}^{(3)} & 0 & 0 & 0 & 0 & \chi_{29}^{(3)} & 0 \\ 0 & 0 & \chi_{33}^{(3)} & 0 & \chi_{35}^{(3)} & 0 & \chi_{37}^{(3)} & 0 & 0 & 0 \end{pmatrix} \begin{pmatrix} E_x^{\omega} E_x^{\omega} E_x^{\omega} \\ E_y^{\omega} E_y^{\omega} E_y^{\omega} \\ E_z^{\omega} E_z^{\omega} E_z^{\omega} \\ 3E_y^{\omega} E_z^{\omega} E_z^{\omega} \\ 3E_y^{\omega} E_y^{\omega} E_z^{\omega} \\ 3E_x^{\omega} E_z^{\omega} E_z^{\omega} \\ 3E_x^{\omega} E_x^{\omega} E_z^{\omega} \\ 3E_x^{\omega} E_y^{\omega} E_y^{\omega} \\ 3E_x^{\omega} E_x^{\omega} E_y^{\omega} \\ 6E_x^{\omega} E_y^{\omega} E_z^{\omega} \end{pmatrix} \tag{2}
$$

where, $\varepsilon_0$ denotes the permittivity of free space. Substituting Equation (1) into Equation (2), we obtain:

$$
\begin{aligned}
\mathrm{P}_{\mathrm{x}}^{3\omega} &= \varepsilon_0 {E_0}^3 [\chi_{11}^{(3)} (\cos(a-\varphi))^3 + 3\chi_{18}^{(3)} \cos(a-\varphi) (\sin(a-\varphi))^2] \\
\mathrm{P}_{\mathrm{y}}^{3\omega} &= \varepsilon_0 {E_0}^3 [\chi_{22}^{(3)} (\sin(a-\varphi))^3 + 3\chi_{29}^{(3)} \sin(a-\varphi) (\cos(a-\varphi))^2] \\
\mathrm{P}_{\mathrm{z}}^{3\omega} &= 0
\end{aligned} \tag{3}
$$

Then, in the laboratory coordinate system (X-Y-Z), the third-order polarization $P^{3\omega}$ can be expressed as:

$$P_X^{3\omega} = \cos\varphi P_x^{3\omega} - \sin\varphi P_y^{3\omega}$$
$$P_Y^{3\omega} = \sin\varphi P_x^{3\omega} + \cos\varphi P_y^{3\omega} \quad (4)$$

The detected THG is given by:

$$I_{THG} = [P_X^{3\omega}]^2 + [P_Y^{3\omega}]^2 \quad (5)$$

By combining Equations (3), (4), and (5), the expression describing the P-THG response at the equator of a starch granule is derived:

$$I_{THG} = A\left[\cos^2(\alpha-\varphi)\left[\left(B\cos^2(\alpha-\varphi) + 3C\sin^2(\alpha-\varphi)\right)\right]^2 + \sin^2(\alpha-\varphi)[\sin^2(\alpha-\varphi) + 3D\cos^2(\alpha-\varphi)]^2\right], \quad (6)$$

where: $A = \varepsilon_0{}^2 E_0{}^6\left(\chi_{22}^{(3)}\right)^2, B = \frac{\chi_{11}^{(3)}}{\chi_{22}^{(3)}}, C = \frac{\chi_{18}^{(3)}}{\chi_{22}^{(3)}}, D = \frac{\chi_{29}^{(3)}}{\chi_{22}^{(3)}}$.

This forms the theoretical basis for interpreting the P-THG signal modulations measured experimentally in corn starch granules. Using the above notations the anisotropy ratio (AR) is defined as follows:

$$AR = \frac{\chi_{11}^{(3)}}{\chi_{22}^{(3)}} = \frac{\chi_{xxxx}^{(3)}}{\chi_{yyyy}^{(3)}}, \quad (7)$$

The AR describes the ratio of the contribution of $\chi_{xxxx}^{(3)}$ relative to the contribution of $\chi_{yyyy}^{(3)}$ to the produced THG signal.

### *2.4. Arithmetic simulations of P-THG modulation*

Figure 3 presents simulated polarization-dependent THG intensity profiles based on Equation (6), which models the angular modulation of the THG signal as a function of excitation polarization. These simulations explore the role of free parameters in shaping the modulation patterns observed in starch granules. In Figure 3a, the anisotropy ratio is set to **AR=1.3**, and the remaining fitting parameters are fixed as A=0.55, C=0.3, and D=0.4. The modulation curves are computed for sample orientations φ ranging from 0° to 180°, in steps of 20°. In Figure 3b, a different set of parameters is chosen: **AR=1.1**, A=0.7, C=0.3, and D=0.5. Importantly, due to the symmetry of the underlying equation, the curves for φ=0° and φ=180°, are identical in both simulation sets. This suggests that the modulation patterns possess a modulo of 180°. These simulations demonstrate how variations in the anisotropy ratio AR and other fitting parameters control the emergence of distinct modulation profiles, offering a robust means of distinguishing structurally different regions within starch granules.

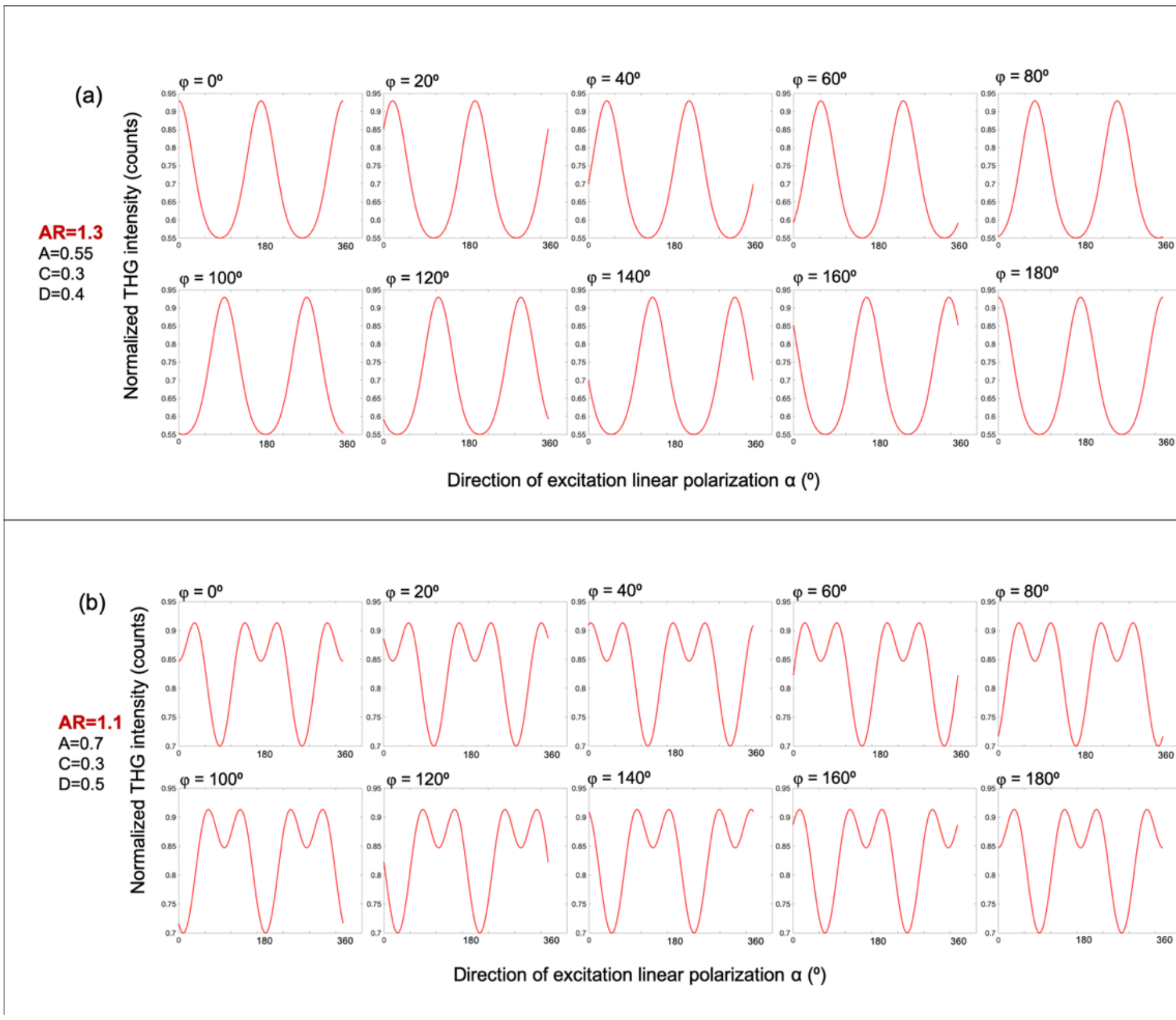

**Figure 3.** Simulated polarization-dependent THG intensity profiles based on Equation (6), illustrating the effect of the anisotropy ratio AR and molecular orientation φ. (a) Simulated P-THG modulation curves for an anisotropy ratio **AR=1.3**, with fixed parameters A=0.5, C=0.3 and D=0.4. The molecular orientation angle φ varies from 0° to 180°, in 20° increments. (b) Corresponding simulations for **AR=1.1**, with A=0.7, C=0.3, and D=0.5. In both panels, the modulation curves for φ=0° and φ=180° are identical, reflecting the modulo of 180° in the calculation of angle φ.

To generate a P-THG modulation curve, the free parameters in Equation (6) are set to the following values: **AR=1.27**, **φ=2°**, A=0.5, C=0.5, and D=0.28, as shown in Figure 4a. Interestingly, a similar modulation profile can be obtained with a different parameter set: **AR=0.79**, **φ=92°**, A=0.9, C=0.2, and D=0.4 (Figure 4b). Likewise, a different modulation pattern is reproduced in Figure 4c using parameters **AR=1.1**, **φ=97°**, A=0.7, C=0.38, and D=0.47, which closely resembles the result in Figure 4d, where the values are **AR=0.9**, **φ=7°**, A=0.85, C=0.43, and D=0.33. These comparisons reveal that the modulation curves remain nearly invariant when φ is shifted by 90° and the anisotropy ratio is simultaneously inverted (i.e., AR→1/AR). This degeneracy suggests that when fitting the experimental data, the iteration algorithm might converge to two different sets of values for the free parameters in Equation (6), that nevertheless describe the same data. To resolve this ambiguity during data fitting, and to obtain comparable values for the free parameters, the criterion AR>1 is imposed to the fitting algorithm.

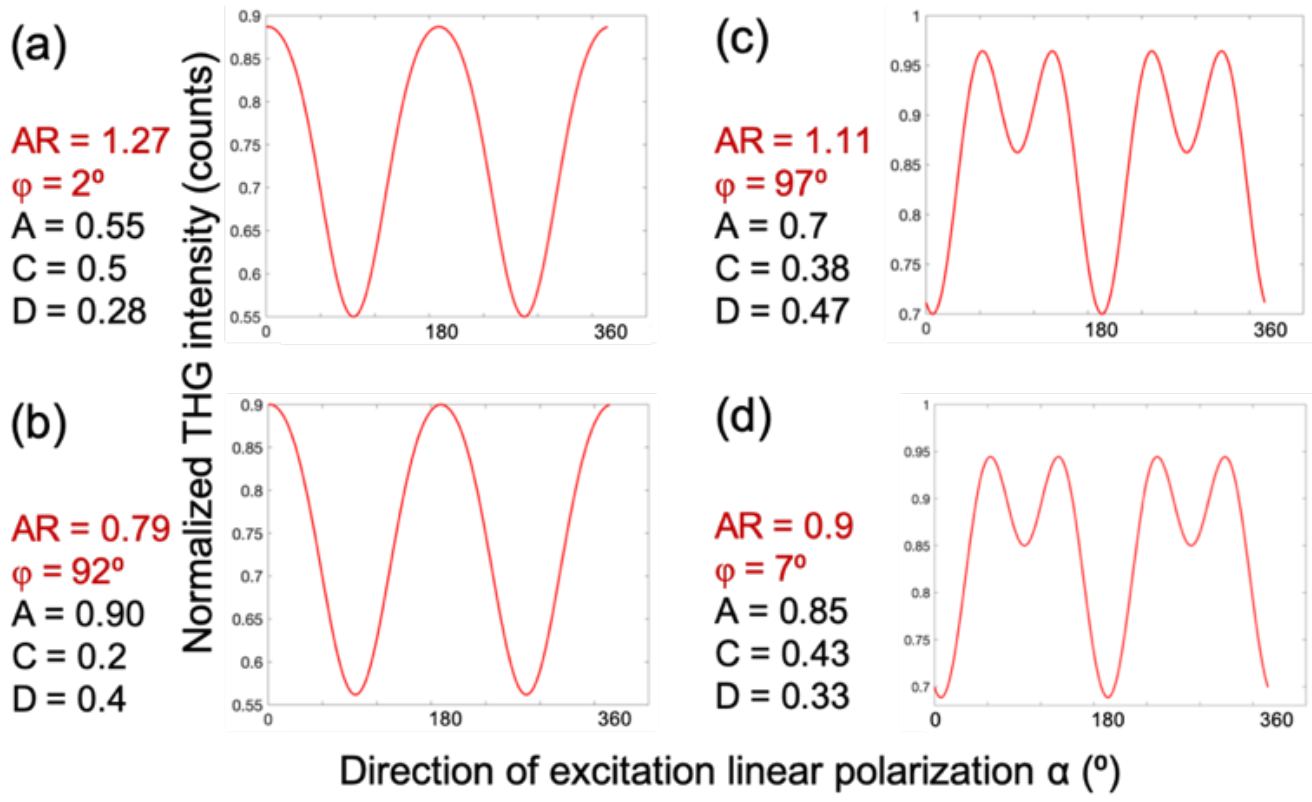

**Figure 4.** Two sets of free parameters in Equation (6) describe the same P-THG data. The P-THG modulation curve for a) molecular angle φ=2° and AR=1.27, is similar to the b) molecular angle φ=92° and AR=0.79. The P-THG modulation curve for c) molecular angle φ=97° and AR=1.11, is similar to the d) molecular angle φ=7° and AR=0.9. Consequently, for the same experimentally retrieved P-THG data, the fitting algorithm might provide two different values for φ and AR. In order to solve this ambiguity and to obtain comparable values of the free parameters, we force the algorithm to provide fitting values with AR>1.

## 3. Results and Discussion

Figure 5a shows the THG signal intensity from corn starch as a function of excitation laser power, plotted on a double logarithmic scale. The observed slope of approximately 3 confirms the cubic power-law dependence, characteristic of a third-order nonlinear optical process like THG [22].

In Figure 5b, a series of THG images, produced by a corn starch granule, recorded while rotating the excitation linear polarization angle $\alpha$ (indicated by the red double-arrow), between 0° and 180° with step 20°, is presented. The images clearly show that the THG signal intensity modulates as a function of the polarization angle $\alpha$. Notably, the image at α=0° is identical to that at α=180°, consistent with the modulo of 180° predicted by our simulations (see Figures 3a and 3b). The THG imaging is performed close to the equator of the granule and both the shell and the core of the granule can be clearly visualized. Importantly, the THG signals originating from the shell of the granule, modulate in a different manner than those originating from the core of the granule. This behavior is more clearly demonstrated in Supporting Video S1, which shows 37 THG images acquired as α varies from 0° to 360° in 10° increments. The modulation patterns observed experimentally match well with our numerical simulations (Figures 3a and 3b), where different parameter sets in Equation (6) yield distinct P-THG modulation profiles.

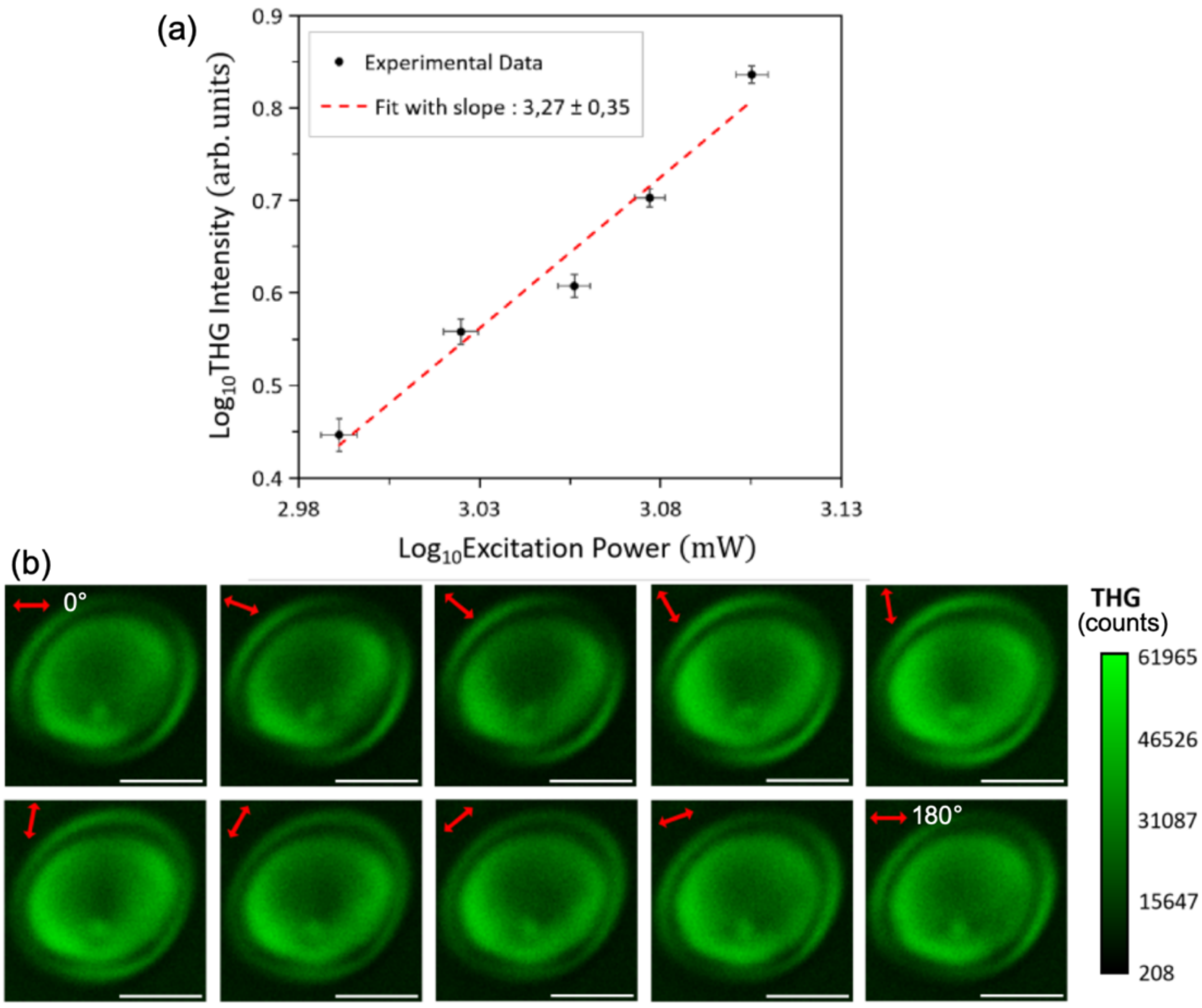

**Figure 5.** Power-law dependency of detected signals and P-THG imaging microscopy of starch and P-THG imaging. a) Log-log plot of the THG intensity from starch as a function of the incident pump power. Black data points with error bars represent the experimental measurements, while the red doted-line shows the linear fit. The slope of approximately 3 confirms the third-harmonic generation process. b) Series of THG images acquired as the excitation linear polarization angle α (indicated by the red double arrow) is rotated from 0° to 180° in 20° increments. It is noted that the image when α is 0° is similar with the image when α is 180°, verifying the modulo of 180° described in the arithmetic simulations. Scale bars: 5μm.

To investigate this further, in Figure 6a five points of interest (POIs) are selected in the shell and five POIs in the core of the granule. In Figure 6b, the P-THG modulations from each POI, are presented (angle $\alpha \in [0°, 360°]$, with a step of 10°). Then, the P-THG experimental data from each POI (Figure 6a) is fitted with Equation (6), using the criterion AR>1 and the P-THG modulation curves shown in Figure 6b are obtained. The values of the free parameters AR, φ, A, C, and D of Equation (6), after fitting the P-THG experimental data with Equation (6), are presented in Table 1. As seen in Table 1, POIs 1-5, which belong to the shell of the granule, exhibited AR values greater than 1.25 whereas the POIs 6-10, which belong to the core of the granule, exhibited AR values less than 1.25.

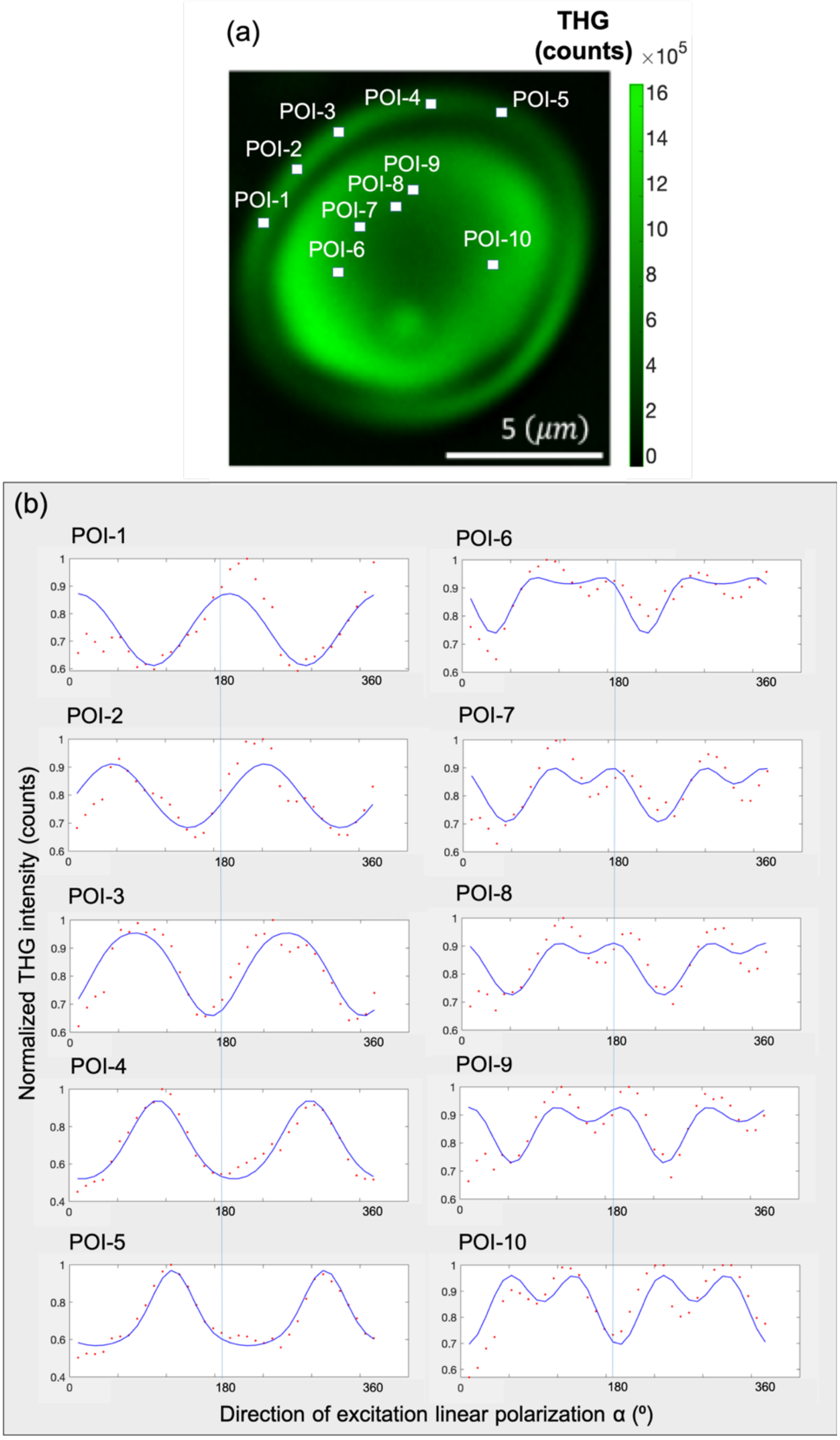

**Figure 6.** Experimental P-THG from selected points-of-interest (POIs) and fitting with the model. a) Composite image formed by summing all the THG images seen in Figure 5b. Five POIs are selected within the shell of the granule and five POIs within its core. b) P-THG modulation curves for each POI, obtained by fitting the experimental data using Equation (6). The POIs 1-5, (shell), present different AR values from POIs 6-10 (core), highlighting the structural anisotropy between the two regions.

**Table 1.** Values of the free parameters AR, φ, A, C, and D, obtained after fitting the experimental P-THG data of the POIs seen in Figure 6a, using Equation (6). The $R^2$ denotes the quality of fitting.

| | **POI-1** | **POI-2** | **POI-3** | **POI-4** | **POI-5** | **POI-6** | **POI-7** | **POI-8** | **POI-9** |
|---|---|---|---|---|---|---|---|---|---|
| **AR** | 1.2±0.03 | 1.16±0.04 | 1.3±0.05 | 1.35±0.06 | 1.27±0.04 | 1.12±0.02 | 1.091±0.03 | 1.097±0.03 | 1.096±0.03 |
| **φ (°)** | 8.8±0.04 | 51.9±0.04 | 75.1±0.03 | 104.9±0.03 | 121.7±0.03 | 126.8±0.02 | 141.7±0.03 | 147.9±0.03 | 152.4±0.03 |
| A | 0.61±0.02 | 0.68±0.03 | 0.56±0.03 | 0.52±0.03 | 0.61±0.02 | 0.87±0.02 | 0.71±0.02 | 0.73±0.02 | 0.73±0.02 |
| C | 0.26±0.02 | 0.37±0.03 | 0.26±0.04 | 0.31±0.05 | 0.12±0.03 | -0.06±0.02 | 0.22±0.02 | 0.24±0.02 | 0.01±0.02 |
| D | 0.48±0.02 | 0.35±0.03 | 0.55±0.03 | 0.42±0.04 | 0.5±0.03 | 0.68±0.02 | 0.54±0.02 | 0.52±0.02 | 0.66±0.02 |
| $R^2$ | 0.59 | 0.65 | 0.88 | 0.92 | 0.93 | 0.64 | 0.5 | 0.47 | 0.43 |

Figure 7a presents a pixel-by-pixel map of the molecular orientation angle φ, obtained by fitting the experimental data from Video S1 using Equation (6). The color scale represents molecular angles from 0° (blue, left side) to 180° (red, right side), illustrating a clockwise evolution of orientation consistent with the rotation of the excitation linear polarization. This continuous color progression across the shell indicates a radial molecular arrangement. Notably, on the right side of the granule, red regions (φ ≈ 180°) are immediately followed by blue regions (φ ≈ 0°). This apparent discontinuity arises from the modulo of 180° in the P-THG signal described in Figure 3a. Since φ and φ + kπ (where k is an integer) produce identical P-THG responses, molecular orientations between 180° and 360° are effectively mapped to the 0°–180° range. For example, φ = 190° is equivalent to φ = 10°, and φ = 360° is equivalent to φ = 180°.

Moreover, in Figure 7a, the molecular orientation in the core appears less uniform, likely reflecting deviations from perfect radial molecular arrangement in the granule. Moreover, it is noted that the evolution of the molecular directions in the core has an ~90° shift with the evolution of the molecular directions in the shell. This is the result of the anisotropic P-THG behavior of the granule between the shell and its core.

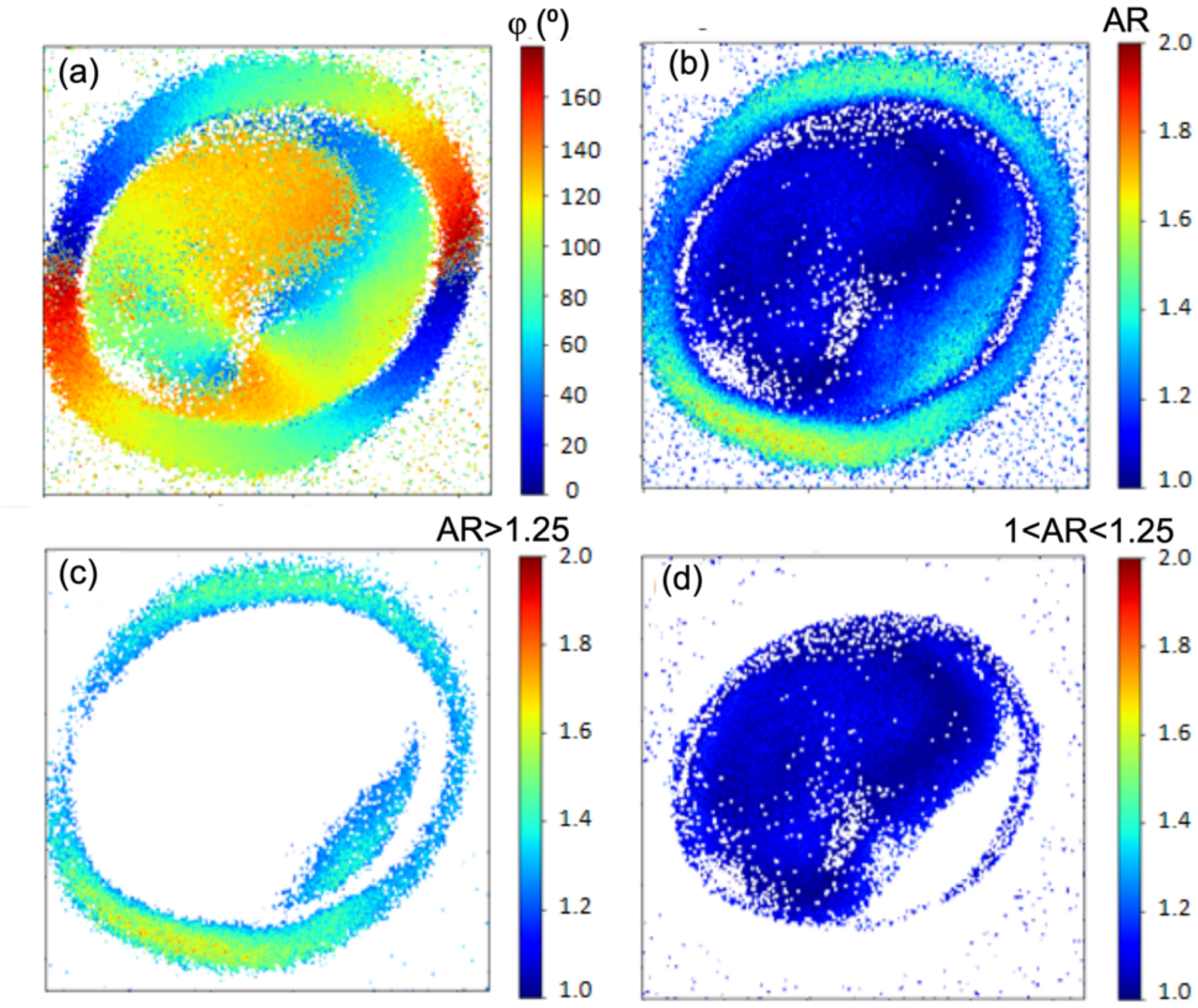

**Figure 7.** Pixel-by-pixel mapping of the molecular orientation angle and the anisotropy ratio. a) Pixel-wise map of the molecular orientation angle φ. The evolution of the colors in the granule shows a radial distribution of the molecular structures. b) Pixel-wise map of the anisotropy-ratio AR. c) Mapping of regions with AR>1.25 highlighting the shell of the granule. (b) Mapping of regions with 1<AR<1.25 exposes the core of the granule**.**

In Figure 7b where a map of the AR parameter is presented, it is noted that the shell predominantly presents AR values bigger than 1.25, while the core of the granule shows mainly AR<1.25. In Figure 7c by selecting pixels with AR>1.25 the shell of the granule is exposed, while in Figure 7d for 1<AR<1.25 the core of the granule is exposed. In the shell (Figure 7c) the mean of the AR values is <AR>=1.37, with standard-deviation σ=0.11, while in the core <AR>=1.14 with σ=0.06. The <AR>=1.37 in the shell implies that the $\chi^{(3)}_{xxxx}$ contributes 1.37 times more in THG, than the $\chi^{(3)}_{yyyy}$. While, the <AR>=1.14 in the core of the granule implies that the $\chi^{(3)}_{xxxx}$ contributes 1.14 times more in THG, than the $\chi^{(3)}_{yyyy}$.

The mean AR values in the shell of the granule are different than the mean AR values in the core. This directly implies that the molecular structure in the shell differs from the molecular structure in the core of the granule. The fact that the “ghost” envelope structure of granules discussed above, has a different structure than the core [7], is clearly reflected in the AR values of polarization-dependent THG microscopy. In Figure S2 and in Video S2, P-THG imaging is presented in another corn starch granule, yielding similar results with those presented above.

## 4. Conclusions

In this study the structural anisotropy between the shell and the core of corn starch granules is probed using P-THG microscopy. By rotating the direction of the excitation linear polarization, the THG signals originating from the shell, modulate in a different manner than the THG signals originating from the core of the granules. A theoretical framework is developed, which describes the P-THG using orthorhombic symmetry, characteristic of corn starch. After pixel-by-pixel fitting of the model and rendering of the retrieved molecular angles, the radial distribution of the molecules is revealed. An anisotropy-ratio, AR=$\chi^{(3)}_{xxxx}/\chi^{(3)}_{yyyy}$, which compares the contribution of the $\chi^{(3)}_{xxxx}$ to the THG signals with the contribution of the $\chi^{(3)}_{yyyy}$, is introduced. The structural anisotropy between the shell and the core is reflected in the AR. By segmenting the values of the AR, the shell and core are discriminated, probing their different molecular architecture. This work provides a label-free methodology to identify structural anisotropy in biological samples and demonstrates how P-THG contrast can reveal different molecular arrangements, beyond what is accessible through intensity-only THG microscopy.

**Acknowledgments:** Maria Kefalogianni acknowledges the project Brainprecision - TAEDR-0535850. This project is carried out within the framework of the National Recovery and Resilience Plan Greece 2.0, funded by the European Union – NextGenerationEU. Sotiris Psilodimitrakopoulos acknowledges the project MAYA – Project number: 014772. This project is carried out within the framework of the National Recovery and Resilience Plan Greece 2.0, funded by the European Union – NextGenerationEU (Implementation body: HFRI).